\documentclass[a4paper]{PoS}

\usepackage{graphicx,amsfonts,amssymb,amsmath,epsfig,amsthm,fancyhdr,eucal,appendix}
\usepackage{caption, subcaption}
\usepackage{fontenc, times, mathptmx}
\usepackage[english]{babel}
\usepackage{simplewick}
\usepackage{cite}
\usepackage[utf8]{inputenc}

\newcommand{\BE}{\begin{equation}}
\newcommand{\EE}{\end{equation}}
\newcommand{\mrm}{\mathrm}
\newcommand{\dd}{\mathrm{d}}

\newcommand{\me}{\mathrm{e}}

\newcommand{\del}{\partial}

\newcommand{\fig}[1]{figure~\ref{#1}}
\newcommand{\Fig}[1]{Figure~\ref{#1}}

\title{Renormalization Group Properties of Scalar Field Theory Using Gradient Flow}

\ShortTitle{Gradient Flow Renormalization Group}

\author{\speaker{Andrea Carosso}$^{\;a}$, Anna Hasenfratz$^a$, Ethan T. Neil$^{a,b}$\\
		$^a$Department of Physics, University of Colorado, Boulder, CO 80309, USA\\
		$^b$RIKEN BNL Research Center, Brookhaven National Laboratory, Upton, NY 11973, USA\\
		E-mail: \email{andrea.carosso@colorado.edu}}

\abstract{Gradient flow has proved useful in the definition and measurement of renormalized quantities on the lattice. Recently, the fact that it suppresses high-modes of the field has been used to construct new, continuous RG transformations both analytically and on the lattice, distinct from the usual blocking techniques in spin models and gauge theories. In this work, we discuss such a formulation for scalar field theory, and we present preliminary numerical results for its application to the determination of critical exponents at the Wilson-Fisher fixed point of three-dimensional $\phi^4$ theory.}

\FullConference{The 36th Annual International Symposium on Lattice Field Theory - LATTICE2018\\
		22-28 July, 2018\\
		Michigan State University, East Lansing, Michigan, USA.}

\begin{document}

\section{Introduction}

Gradient flow (GF) is by now a popular tool for the definition and computation of renormalized quantities on the lattice \cite{Luscher:2010iy}. It has been used to set the scale in lattice QCD and to compute running couplings, along with providing improved ways for computing beta functions from step-scaling \cite{Hasenfratz:2015xpa}. Essential to all these applications is the characterstic smoothing property implied by its general form as a heat equation. At this point, one cannot help but note a resemblance with the most prototypical of RG transformations -- spin-blocking. There, one defines blocked spins as local averages of bare spins. Under GF, the flowed field at a point is a weighted average of the initial one, with flow time $t$ parametrizing the effective width.

Although GF provides the characterstic smoothing one wants of an RG tranformation, it is clear that it is not enough. RG transformations involve an explicit rescaling of the degrees of freedom in order to allow for the existence of fixed points. By supplementing GF with such a rescaling, one may reasonably expect to then have a viable RG transformation. Such an approach \cite{Carosso:2018} has already been applied in an $SU(3)$, $N_f = 12$ system to obtain the fermion mass anomalous dimension, consistent with several previous results \cite{Gracey:2018oym,Cheng:2013bca}, and has led to the first lattice determination of the baryon anomalous dimension. Analytic aspects of GF in relation to RG have been discussed  in recent work \cite{Monahan:2015lha,Makino:2018rys}.

\begin{figure}
\centering
\begin{minipage}{.45\textwidth}
  \centering
\includegraphics[width=0.7\textwidth]{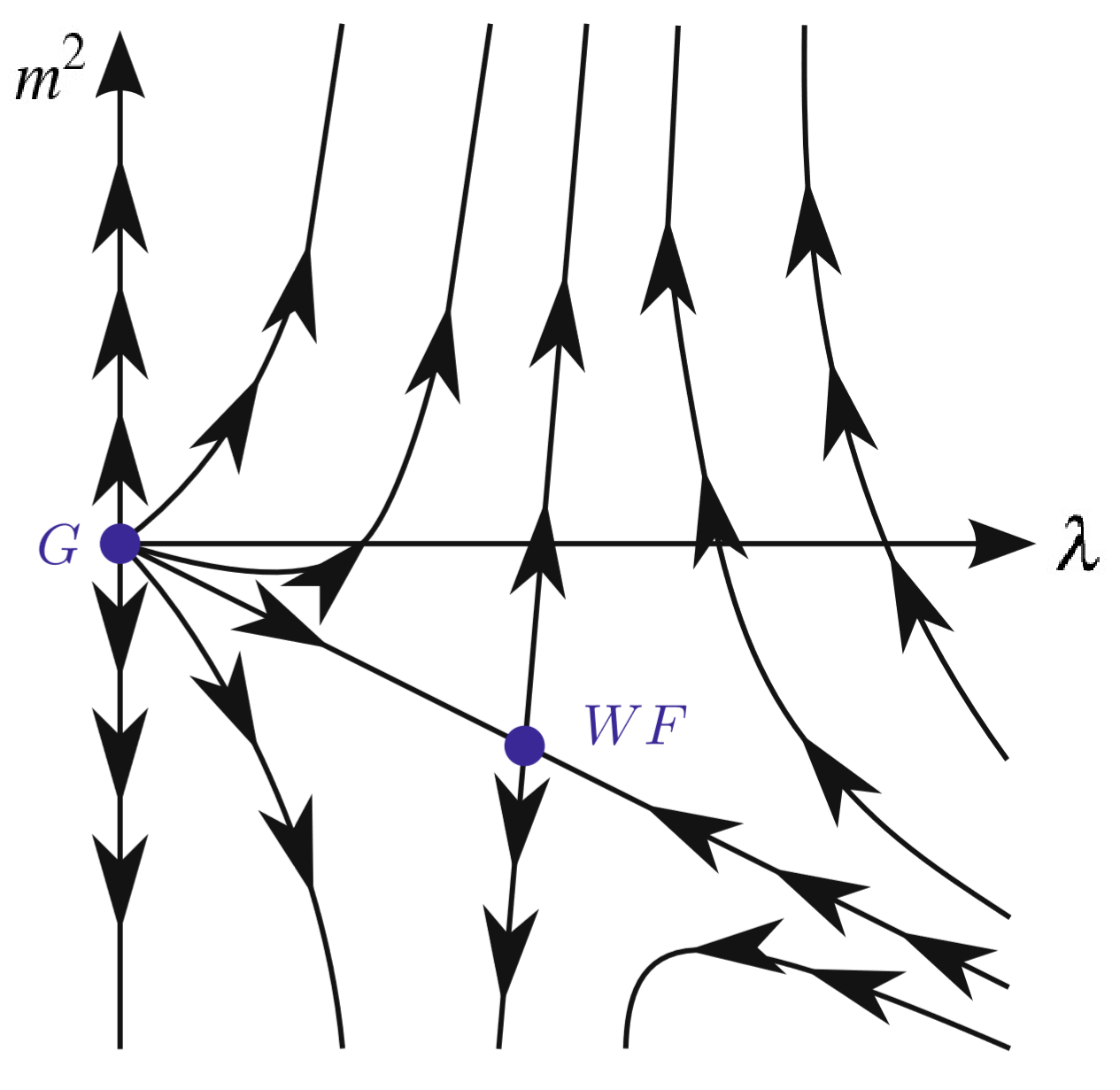}
\caption{\small{Projection to the $\lambda-m^2$ plane of the RG flow near the Wilson-Fisher fixed point. Adapted from \cite{Kopietz:2010zz}. \label{fig:WFFP}}}
  \label{fig:test1}
\end{minipage}\hfill
\begin{minipage}{.45\textwidth}
  \centering
\includegraphics[width=1.1\textwidth]{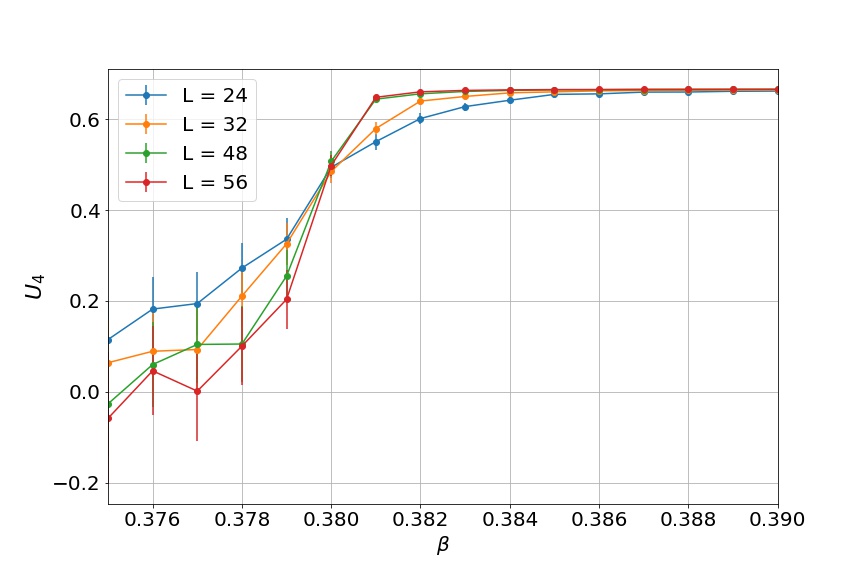}
\caption{\small{The Binder cumulant takes a universal value at the critical coupling $\beta_c$ for $L \to \infty$.} \label{fig:binder}}
  \label{fig:test2}
\end{minipage}
\end{figure}

As a check on the general viability of this transformation, one can test it on a well-known system, such as scalar $\phi^4$ theory in three dimensions, where one expects to measure the critical exponents of the Wilson-Fisher fixed point (WFFP). In this contribution we report on the initial results of such an attempt. The infinite volume lattice action is given by
\BE \label{phi4lat}
S(\varphi) = \sum_{n\in \mathbb{Z}^d} \Big[ - \beta \sum_{\mu = 1}^d  \varphi(n)\varphi(n+\mu) + \varphi^2(n) + \lambda(\varphi^2(n) - 1)^2 - \lambda \Big],
\EE
where all the quantites are dimensionless, and are related to the dimensionful quantities $m^2, \; g$ in a straight-forward way. For $d=3$, the system possesses an infrared conformal fixed point, the WFFP. It is characterized by one relevant parameter, the mass-squared $m^2$, which is related to $\beta$. The rest of the couplings, such as the quartic coupling $\lambda = g_4$, or the six-point coupling $g_6$, and so on, are all irrelevant at the fixed point. \Fig{fig:WFFP} depicts the projection of the RG flow to the $m^2-\lambda$ plane.

Close to the fixed point, scaling operators transform under RG as $O_i' = b^{\Delta_i} O_i$, where $\Delta_i$ is the scaling dimension of the operator. It is customarily written as $\Delta_i = d_i + \gamma_i$, where $d_i$ is the canonical, and $\gamma_i$ is the anomalous dimension of the operator. If $O_i$ is in the action, its coupling, $u_i$, scales with RG eigenvalue $y_i = d - \Delta_i$. Such $O_i$ will generally be linear combinations of familiar action operators, such as those appearing in eq. \ref{phi4lat}. Relevant operators have $y_i > 0$, while irrelevant operators have $y_i < 0$. Operators with $y_i = 0$ are marginal.

The field $\phi$ has $d_\phi = 1/2$ for $d=3$. The anomalous dimension $\gamma_\phi$ of $\phi$ is related to the critical exponent $\eta= 0.0358(4)$ \cite{Hasenbusch:1999mw} by $\gamma_\phi = \eta/2 \approx 0.018$, hence $\Delta_\phi \approx 0.518$. The operator $\phi^2$ has anomalous dimension $\gamma_{\phi^2} = d - 1 - 1/\nu \approx 0.412$, where $\nu = 0.6296(3)$, so that $\Delta_{\phi^2} \approx 1.412$. Lastly, $\phi^4$ has $\gamma_{\phi^4} =d-2+\omega \approx 1.845$, where $\omega = 0.845(10)$, so that $\Delta_{\phi^4} \approx 3.845$. We find it worthwhile to point out that these operators correspond to RG eigenvalues $y_\phi \approx 2.482,\; y_{\phi^2} \approx 1.588, \; y_{\phi^4} \approx -0.845$, which implies that $\phi$ and $\phi^2$ are relevant, while $\phi^4$ is irrelevant. 

\section{The Binder Cumulant}

To study the critical properties of a system, one must tune the bare parameters of the simulation to be close to the critical surface. If the initial parameters lie slightly off this surface, then repeated RG transformations at first might approach the fixed point, but eventually veer away past it in the relevant directions. When there is one relevant parameter, say $m^2$, then this means tuning $m^2$ to be close to a particular value $m^2_\mrm{c}$ in order to approach the fixed point.

In terms of simulation parameters, it is the neighbor coupling $\beta$ that must be tuned. The most practical method for doing so is to use the Binder cumulant \cite{Binder:1981sa}. It is defined by
\BE
U_4(\beta,L) = \frac{3}{2} - \frac{\langle M^4 \rangle}{2 \langle M^2 \rangle^2},
\EE
where $M = V^{-1}\sum_n \varphi(n)$ is the magnetization. At the critical point and in the infinite volume limit, $U_4(\beta_c, \infty) = U_4^*$ takes a universal value. The leading deviations from its fixed-point value due to finite volume may be parametrized by the leading irrelevant exponent $\omega$:
\BE
U_4(\beta_c,L) = U_4^* + c_1L^{-\omega}.
\EE
Thus, for fixed $\lambda$, one measures $U_4$ on various volumes and at many $\beta$ values, and extrapolates to $L = \infty$ at each $\beta$, searching for the extrapolation which matches most closely the universal value $U_4^*$ \cite{Amit:1984ms}. Estimates for this value can be found in \cite{Hasenbusch:1999mw}, where the best fit predicts $0.69819(12)$. 

In our simulation, we have chosen $\lambda = 1.00$ and searched for the critical value $\beta_c$. It was noted in \cite{Hasenbusch:1999mw} that $\lambda = 1.1$ exhibits a minimum of corrections to scaling. We therefore expect that $\lambda = 1.00$ does not exhibit dramatic scaling violations either, though further testing must be done. We found that $\beta = 0.38$ yields an estimate of $U_4^* = 0.774(17)$, which we consider close \emph{enough} to the critical surface, given the encouraging results for exponents discussed later. Future work will involve a determination of how much tuning of $\beta$ is required. Simulation details will be given in section 4.

\section{From Gradient Flow to Gradient Flow Renormalization Group}

Here we derive a relation between GF and RG, in the same spirit as in Ref. \cite{Carosso:2018}. Begin with the usual definition of a block-spin that one commonly finds, for example, in Cardy \cite{Cardy:1996xt}:
\BE
\varphi_b(n_b) = \frac{b^{\Delta_\phi}}{b^d} \sum_\varepsilon \varphi(n+\varepsilon),
\EE
where $n+\varepsilon$ are the sites in the block ($\varphi$ is dimensionless). The number $n_b = n/b$ denotes the lattice site on the reduced lattice, while $n$ belongs to the un-blocked lattice. The presence of $\Delta_\phi$ follows from the requirement that the RG transformation should have a fixed point.

Now, the free gradient flow equation is a heat equation, $\del_t \phi_t(n) = (\Delta \phi_t)(n),$ with $\phi_0(n) = \varphi(n)$, where $\Delta$ is the lattice Laplacian. The solution in the continuum limit is given by action of the heat kernel $K_t$ on the initial field configuration $\varphi$:
\BE
\phi_t(n) = (K_t \varphi)(n) = \int_{\mathbb{R}^d} \dd^d \varepsilon  \; \frac{\me^{-|\varepsilon|^2/4t}}{(4\pi t)^{d/2}} \; \varphi(n+\varepsilon).
\EE
The flowed field is effectively a local average of the initial field within a radius $\approx\sqrt{8t}$. If we identify $b \propto \sqrt{t}$, we observe that the definition $\phi_b(n/b) := b^{\Delta_\phi} \phi_t(n)$ has the characteristic features of a block-spin that we want for an RG transformation. Since the blocking factor need no longer be integer, the argument $n_b = n/b$ does not refer to a real lattice site, in general. We will not need an exact description of the blocked lattice for our result, however. 

Next, let us recall the RG scaling relations of correlators of scaling operators \cite{Cardy:1996xt, DelDebbio:2010ze}:
\BE
\langle O_b(n/b) O_b(0) \rangle \approx b^{2\Delta_O} \langle O(n) O(0) \rangle.
\EE
Define a blocked operator in terms of $\phi_b$ as $O_b(n/b) := O(\phi_b; n)$. Abbreviating $O(\phi_t; n) = O_t(n)$, then the flowed operators satisfy
\BE
\langle O_t(n) O_t(0) \rangle \approx b^{2(\Delta_O - n_O \Delta_\phi)} \langle O(n) O(0) \rangle,
\EE
where $n_O$ is the degree of the polynomial $O(\phi;n)$ in $\phi$. 

If we restrict our attention to operators that have no derivatives, then their scaling dimension may be written as $\Delta_O = n_O d_\phi + \gamma_O$. Since $\Delta_\phi = d_\phi + \gamma_\phi$ and $b \propto \sqrt{t}$, we have derived a ratio
\BE
\frac{\langle O_t(n) O_t(0) \rangle}{\langle O(n) O(0) \rangle} = b^{2(\gamma_O - n_O \gamma_\phi)}
\EE
for flowed correlators close to the fixed point. We have glossed over the subtlety that to get close to the fixed point, the scale change $b$, and therefore the flow time, must be large enough. Comparing two flowtimes $t' > t$ then implies the scaling formula
\BE \label{ratio}
R_O(t',t) := \frac{\langle O_{t'}(n) O_{t'}(0) \rangle}{\langle O_t(n) O_t(0) \rangle} = \Big(\frac{t'}{t} \Big)^{\gamma_O - n_O \gamma_\phi}.
\EE
Notice that we do not need to know the proportionality factor that relates $b$ to $\sqrt{t}$. One may therefore measure the difference $\gamma_O - n_O \gamma_\phi$ numerically; if we input a known value for $\gamma_\phi$, then we may predict $\gamma_O$. On the other hand, if it is known that $\gamma_O = 0$, then we may measure $\gamma_\phi$. Lastly, we stress  that this formula holds only if $O$ is a scaling operator, which is in general a linear combination of action operators.

\section{Results}

Before turning to the results of application of the ratio formula, let us briefly describe the simulation details. We simulated the lattice theory determined by eq. \ref{phi4lat} on hypercubic lattices of sizes $L = 24, \; 36, \; 48$. Our  update sweep consisted of one Wolff cluster update and one Metropolis update for the length of the spins, with maximum update length $d_\mrm{rad} = 0.65$; see \cite{Hasenbusch:1999mw} for a description of the radial update. The acceptance ratio for the radial update was $0.88$. We estimate an autocorrelation time $\tau_\mrm{int} \approx 40$, which is dominated by the radial update; measurements were taken every 5 sweeps. Future work will optimize $d_\mrm{rad}$ to yield higher statistics per  computation time.

In testing the ratio formula, eq. \ref{ratio}, we have chosen $t' = t + 0.25$, in which case we write $R_O(t,n)$, where $t$ is the earlier time, and $n$ is the lattice separation. We first applied the ratio formula to the case $O = \phi$, but one should note that this operator cannot be used to measure the field anomalous dimension $\gamma_\phi$, since in this case the formula reads $R_\phi(t,n) =1$. Nevertheless, one can check for the validity of this relation. Indeed, as illustrated in Figure 3, we verified that the ratio $R_\phi(t,n)$ approaches 1 at distances $n > 2 \sqrt{8t}$, and that the corresponding ``exponent'' approaches zero as the volume increases. The relevant distance here is $2\sqrt{8t}$ because two spins, smeared by $\sqrt{8t}$, will then begin to overlap.

\begin{figure}
\centering
\begin{minipage}{.45\textwidth}
  \centering
  \includegraphics[width=1.1\linewidth]{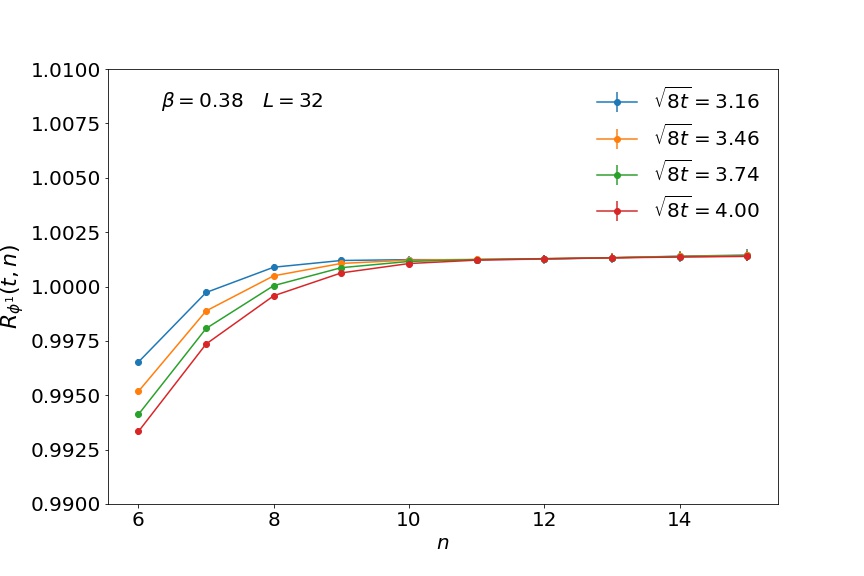}
  \captionof{figure}{\small{$R_\phi$ at several flow times, as a function of spin separation.}}
  \label{fig:test1}
\end{minipage}\hfill
\begin{minipage}{.45\textwidth}
  \centering
  \includegraphics[width=1.1\linewidth]{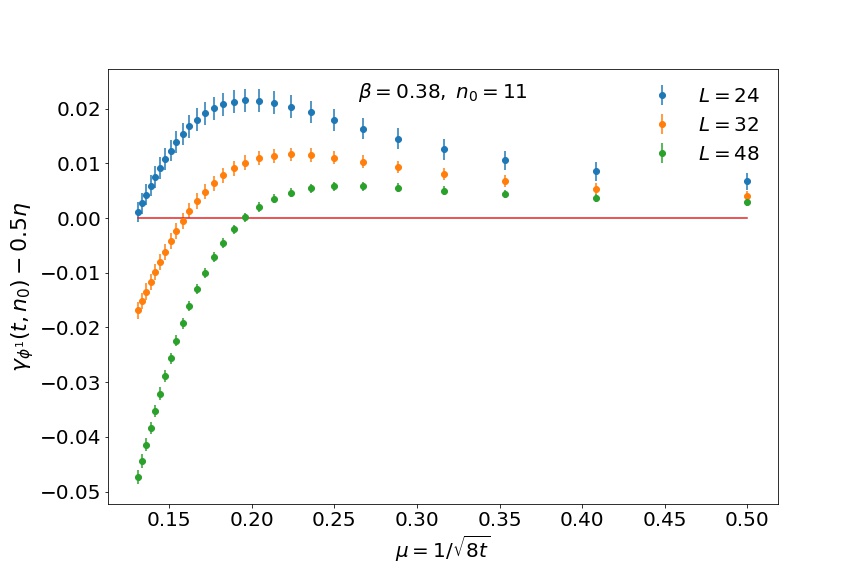}
  \captionof{figure}{ \small{The ``exponent'' decreases with increasing $L$. A line at 0 is displayed for reference. }}
  \label{fig:test2}
\end{minipage}
\end{figure}

\begin{figure}
\centering
\begin{minipage}{.45\textwidth}
  \centering
  \includegraphics[width=1.2\linewidth]{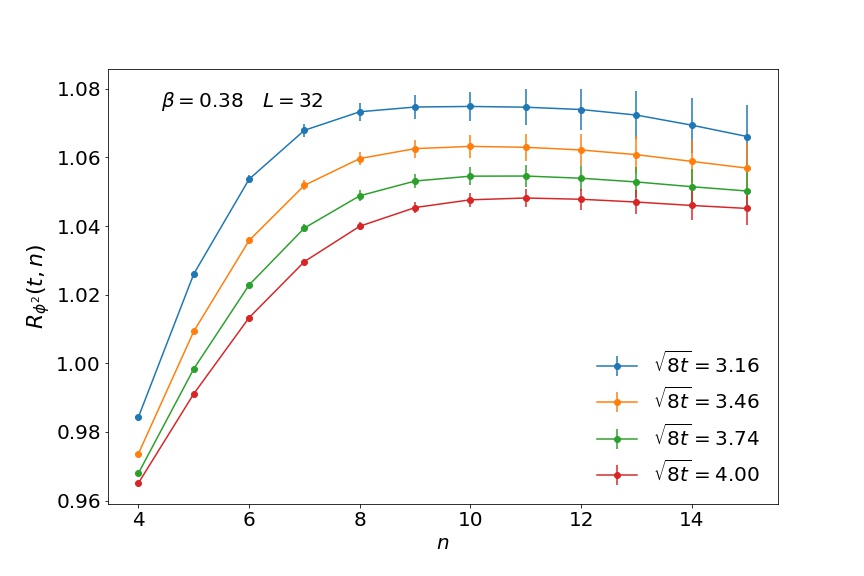}
  \captionof{figure}{\small{$R_{\phi^2}$ as a function of operator separation.}}
  \label{fig:phi2ratio}
\end{minipage}\hfill
\begin{minipage}{.45\textwidth}
  \centering
  \includegraphics[width=1.2\linewidth]{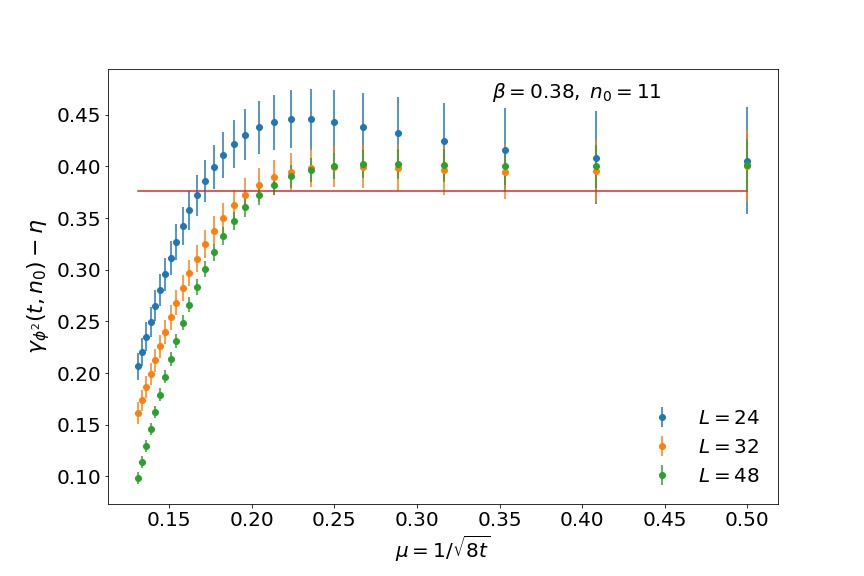}
  \captionof{figure}{\small{$\gamma_{\phi^2}-\eta$. A line is displayed at the known value 0.376. }}
  \label{fig:phi2exp}
\end{minipage}
\end{figure}

\begin{figure}
\centering
\begin{minipage}{.45\textwidth}
  \centering
\includegraphics[width=1.2\textwidth]{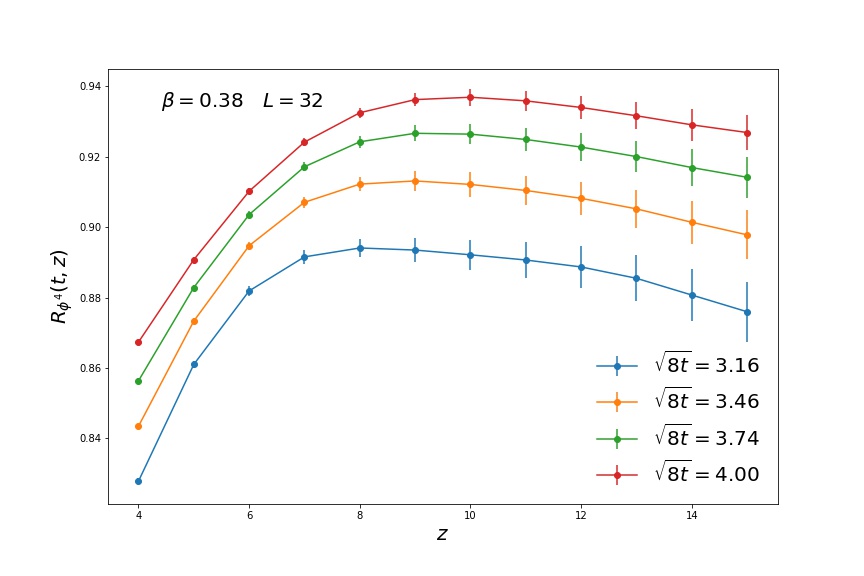}
\caption{\small{Ratios of $O=\phi^4$ correlations show notable slopes at large distance. \label{fig:binder}}}
  \label{fig:phi4ratio}
\end{minipage}\hfill
\begin{minipage}{.45\textwidth}
  \centering
\includegraphics[width=1.2\textwidth]{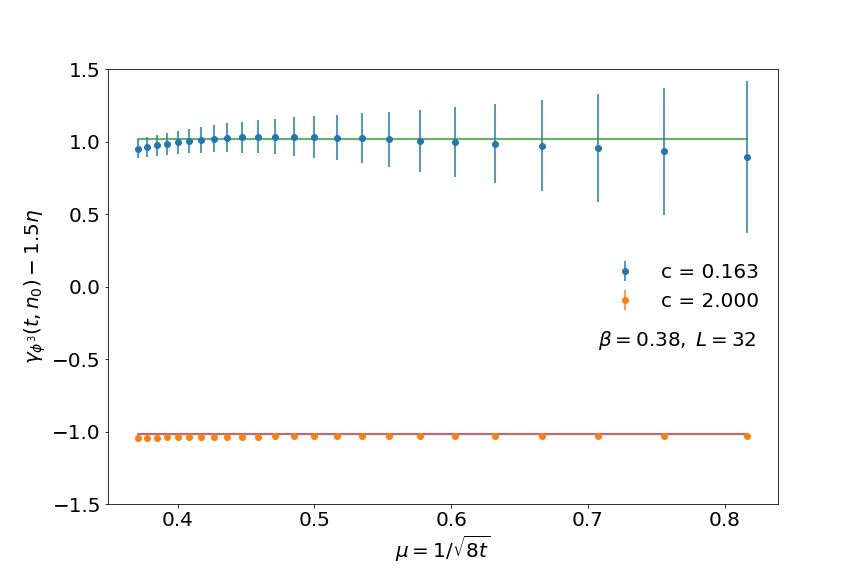}
\caption{\small{The value $ c=0.163$ reveals a plateau corresponding to the exponent $\gamma_{\phi^3}$.}}
  \label{fig:phi3exp}
\end{minipage}
\end{figure}

Turning now to $O = \phi^2$, one can measure $\gamma_{\phi^2} - 2 \gamma_\phi$, which should be about $0.376$. In \fig{fig:phi2exp} we plot the measured value at several flowtimes. It is approximately constant over a range of flowtimes, with a value around $0.4$, slightly higher than expected. Looking at the correlator ratio, we see that a plateau forms at large distances, except there seems to be a slight slope. Here we recall that we do not expect these simple action-operators to be equal to scaling operators whose ratio formula leads to a pure anomalous dimension. It is therefore surprising that this quantity yields a relatively stable value, in rough agreement with the known exponent. For higher-powers of $\phi$, the correlator ratios show significant sloping, suggesting that the corresponding scaling operators have large contributions from other operators. The example of $\phi^4$ is shown in \fig{fig:phi4ratio}.

To better approach the problem, one can form a linear combination of operators and vary their relative coefficients, searching for a plateau. We attempted a simple form
\BE
O^{(3)}_t(x) = \phi_t^3(x) + c \; t^{-\Delta_\phi} \phi_t(x),
\EE
where $c$ is a parameter that we scan over. The presence of $t^{-\Delta_\phi}$ is necessitated by our definition of $O_b$. We input the known value of $\eta$ by hand, since we seek to measure $\gamma_{\phi^3}$. First note that for very large $c$ values, the ratio should simply reproduce the result from $O = \phi$, except shifted downward by $2\Delta_\phi = 1.036$, as illustrated in \fig{fig:phi3exp}. Additionally, there was a narrow range of $c$ values around $c = 0.163$ that located a distinct plateau, corresponding to an anomalous dimension close to $1$. While the plateau for large $c$ is quite steady, we observed that the plateau for $c=0.163$ is not nearly as flat. This value for $\gamma_{\phi^3}$ is consistent with the prediction $\Delta_{\phi^3} = 2 + \Delta_{\phi}$ from \cite{Rychkov:2015naa}.

\section{Discussion}

We have shown that the leading relevant anomalous dimension $\gamma_{\phi^2}$ can be measured relatively easily using a correlator ratio formula that follows from supplementing GF by a rescaling. The dimensions of higher operators are more difficult, due possibly to (1) not fine-tuning $\beta$ enough, (2) large errors in their long-distance correlators, and (3) the presence of scaling from not looking at pure scaling operators.  The variational method presented above for measuring the $\gamma_{\phi^3}$ exponent attempts to circumvent the last issue, but it is admittedly somewhat brute-forced. In future work, together with higher statistics and better tuning, we will implement a more systematic approach by forming a basis of operators $\{ S_i \}$ and measuring the mixed correlations $C_{ij}(n) = \langle S_i(n) S_j(0) \rangle_c$ at finite flow time. By diagonalizing the matrix of correlators one can find estimates for the correlations of scaling operators, to which the ratio formula should apply.

\paragraph{Acknowledgements -} This work was supported by the U.S. Department of Energy under grant DE-SC0010005. Brookhaven National Laboratory is supported by the U.S. Department of Energy under contract DE-SC0012704.

\bibliographystyle{JHEP}
\bibliography{RG_GF}

\providecommand{\href}[2]{#2}\begingroup\raggedright\begin{thebibliography}{10}

\bibitem{Luscher:2010iy}
M.~Lüscher, \emph{{Properties and uses of the Wilson flow in lattice QCD}},
  \href{https://doi.org/10.1007/JHEP08(2010)071,
  10.1007/JHEP03(2014)092}{\emph{JHEP} {\bfseries 08} (2010) 071}
  [\href{https://arxiv.org/abs/1006.4518}{{\ttfamily 1006.4518}}].

\bibitem{Hasenfratz:2015xpa}
A.~Hasenfratz, \emph{{Improved gradient flow for step scaling function and
  scale setting}}, \href{https://doi.org/10.22323/1.214.0257}{\emph{PoS}
  {\bfseries LATTICE2014} (2015) 257}
  [\href{https://arxiv.org/abs/1501.07848}{{\ttfamily 1501.07848}}].

\bibitem{Carosso:2018}
A.~Carosso, A.~Hasenfratz and E.~Neil, \emph{{Nonperturbative renormalization
  of operators in near-conformal systems using gradient flow}}, {\emph{Phys.
  Rev. Lett. (to appear)} }.

\bibitem{Gracey:2018oym}
J.~A. Gracey, T.~A. Ryttov and R.~Shrock, \emph{{Scheme-Independent
  Calculations of Anomalous Dimensions of Baryon Operators in Conformal Field
  Theories}}, \href{https://doi.org/10.1103/PhysRevD.97.116018}{\emph{Phys.
  Rev.} {\bfseries D97} (2018) 116018}
  [\href{https://arxiv.org/abs/1805.02729}{{\ttfamily 1805.02729}}].

\bibitem{Cheng:2013bca}
A.~Cheng, A.~Hasenfratz, G.~Petropoulos and D.~Schaich, \emph{{Determining the
  mass anomalous dimension through the eigenmodes of Dirac operator}},
  \href{https://doi.org/10.22323/1.187.0088}{\emph{PoS} {\bfseries LATTICE2013}
  (2014) 088} [\href{https://arxiv.org/abs/1311.1287}{{\ttfamily 1311.1287}}].

\bibitem{Monahan:2015lha}
C.~Monahan and K.~Orginos, \emph{{Locally smeared operator product expansions
  in scalar field theory}},
  \href{https://doi.org/10.1103/PhysRevD.91.074513}{\emph{Phys. Rev.}
  {\bfseries D91} (2015) 074513}
  [\href{https://arxiv.org/abs/1501.05348}{{\ttfamily 1501.05348}}].

\bibitem{Makino:2018rys}
H.~Makino, O.~Morikawa and H.~Suzuki, \emph{{Gradient flow and the Wilsonian
  renormalization group flow}},
  \href{https://arxiv.org/abs/1802.07897}{{\ttfamily 1802.07897}}.

\bibitem{Hasenbusch:1999mw}
M.~Hasenbusch, \emph{{A Monte Carlo study of leading order scaling corrections
  of $\phi^4$ theory on a three-dimensional lattice}},
  \href{https://doi.org/10.1088/0305-4470/32/26/304}{\emph{J. Phys.} {\bfseries
  A32} (1999) 4851} [\href{https://arxiv.org/abs/hep-lat/9902026}{{\ttfamily
  hep-lat/9902026}}].

\bibitem{Kopietz:2010zz}
P.~Kopietz, L.~Bartosch and F.~Schütz, \emph{{Introduction to the functional
  renormalization group}},
  \href{https://doi.org/10.1007/978-3-642-05094-7}{\emph{Lect. Notes Phys.}
  {\bfseries 798} (2010) 1}.

\bibitem{Binder:1981sa}
K.~Binder, \emph{{Finite size scaling analysis of Ising model block
  distribution functions}}, \href{https://doi.org/10.1007/BF01293604}{\emph{Z.
  Phys.} {\bfseries B43} (1981) 119}.

\bibitem{Amit:1984ms}
D.~J. Amit and V.~Mart\'{i}n-Mayor, \emph{{Field Theory, The Renormalization
  Group, and Critical Phenomena, 3rd Edition}}. Singapore, World Scientific,
  2005.

\bibitem{Cardy:1996xt}
J.~L. Cardy, \emph{{Scaling and renormalization in statistical physics}}.
  Cambridge, UK: Univ. Pr. (1996) 238 p. (Cambridge lecture notes in physics:
  3), 1996.

\bibitem{DelDebbio:2010ze}
L.~Del~Debbio and R.~Zwicky, \emph{{Hyperscaling relations in mass-deformed
  conformal gauge theories}},
  \href{https://doi.org/10.1103/PhysRevD.82.014502}{\emph{Phys. Rev.}
  {\bfseries D82} (2010) 014502}
  [\href{https://arxiv.org/abs/1005.2371}{{\ttfamily 1005.2371}}].

\bibitem{Rychkov:2015naa}
S.~Rychkov and Z.~M. Tan, \emph{{The $\epsilon$-expansion from conformal field
  theory}}, \href{https://doi.org/10.1088/1751-8113/48/29/29FT01}{\emph{J.
  Phys.} {\bfseries A48} (2015) 29FT01}
  [\href{https://arxiv.org/abs/1505.00963}{{\ttfamily 1505.00963}}].

\end{thebibliography}\endgroup



\end{document}